# Sharp Rise in Cosmic Ray Irradiation of Organisms on Earth Caused by a Nearby SN Shockwave Passage

A.A. Shchepkin[1†], G.I. Vasilyev[2], V.M. Ostryakov[1] and A.K. Pavlov[2*]

[1]Peter the Great St. Petersburg Polytechnic University Polytechnicheskaya st., 29, 195251, Saint-Petersburg, Russia;

[2]Ioffe Institute of Russian Academy of Sciences, Polytechnicheskaya st., 26, 194021, Saint-Petersburg, Russia

E-mail:   [†]alexander.shchepkin5@gmail.com, [*]anatoli.pavlov@mail.ioffe.ru

Abstract: The work considers the modeling of nearby supernova (SN) effects on Earth's biosphere via cosmic rays (CRs) accelerated by shockwaves. The rise of the radiation background on Earth resulted from the external irradiation by CR high energy particles and internal radiation in organisms by the decay of cosmogenic $^{14}C$ is evaluated. We have taken into account that the CR flux near Earth goes up steeply when the shockwave crosses the Solar System, while in previous works the CR transport was considered as purely diffusive. Our simulations demonstrate a high rise of the external ionization of the environments at Earth's surface by atmospheric cascade particles that penetrate the first 70-100 m of water depth. Also, the cosmogenic $^{14}C$ decay is able to irradiate the entire biosphere and deep ocean organisms. We analyzed the probable increase in mutation rate and estimated the distance between Earth and an SN, where the lethal effects of irradiation are possible. Our simulations demonstrate that, for SN energy of around $10^{51}$ erg, the lethal distance could be ~ 18 pc.

## 1     Introduction

According to $^{60}Fe$ concentration measurements in the deep-sea manganese crust, at least two SNe explosions have occurred at a distance of about 50-100 pc from Earth during the last 10 My (Knie *et al*., 2004; Fry *et al*., 2016; Wallner *et al*., 2021). This may have been caused by the Sco-Cen OB association activity near the Solar System (Benitez *et al*., 2002; Breitschwerdt *et al*., 2016). The average evolution time of giant stars is ~ 100 My (My - million years), which is much less than the evolution time of Sun-like stars (~ $10^4$ My).

Therefore, the rate of SNe explosions near Earth in the past was greater than the current rate. The shockwaves generated by a series of SNe have formed a Local Bubble, and since $^{60}$Fe isotope has been detected on Earth, this bubble passed through the Solar System, and at this moment, our planet is inside of it (Frisch *et al*., 2009; Zucker *et al*., 2022; Breitschwerdt *et al*., 2016). According to acceleration mechanisms, there is a big jump in the CR intensity at the shockwave and beyond. So, the local CR flux is increased when the SN shock has crossed the Solar System. Then it remains almost constant during the period from hundreds to hundreds of thousands of years depending on the SN kinetic energy $E_{SN}$ and on the distance $L$ between Earth and an SN. Such a leap in the high energy CR flux must enhance the radiation background on Earth by a few orders and must lead to the rise in the rate of mutations in organisms, which makes their mass extinctions possible. The highest registered $^{60}$Fe flux corresponds to the SN event at 2.6 My ago and coincides with the Pliocene-Pleistocene (PP) boundary, when about 36% of marine megafauna went extinct and new kinds of hominids appeared including our kind *homo* (Pimiento *et al*., 2017; deMenocal, 2004; Stanley, 1986). It should be noted that the geomagnetic reversal Gauss-Matuyama occured during the period of $2.4 - 2.9$ My ago. This means that Earth's magnetic field at that time was substantially lower than the current value (Deino *et al*., 2006) that strengthens the CR impact.

One of the first attempts to study the terrestrial effects of a nearby SN was undertaken by Ruderman (1974) and Whitten *et al*. (1976) to estimate the biological effect of stratospheric ozone depletion due to UV, gamma, and cosmic rays from a nearby SN. Gehrels *et al*. (2003) were the first to have investigated this problem with model calculations. Their estimation of maximal distance where lethal effects are possible is $L_{le} = 8$ pc for SNe with $E_{SN} = 10^{51}$ erg. In the recent work of Brunton *et al*. (2023), the lethal effects of an X-ray luminous SN on Earth's ozone depletion were studied. These found that the most X-ray luminous SNe with total X-ray energy $E_X \sim 10^{49}$ erg can lead to lethal consequences on Earth at the distance of $L_{le} = 20 - 50$ pc. Melott *et al*. (2017) and Thomas and Yelland (2023) calculated the muon flux from air showers generated by CRs from a nearby SN, and Melott *et al*. (2019) investigated the possibility of marine megafaunal extinctions caused by the increase in muon flux in oceanic water. These found that, for an SN at $L = 50$ pc, the radiation dose contributed by muon flux in oceanic water for depths up to 1 km could exceed the present radiation dose from all terrestrial sources. This increase could influence the processes of extinctions of marine megafauna.

However, the shockwave structure of CR sources has not been considered until now. The diffusive CR transport model is not valid when the SN shock crosses the Solar system and the jump in the CR intensity appears. Moreover, purely diffusive transport of the cosmic rays is not fully correct because the diffusion equation accepts solutions that violate the causality principle. Also, in these works only muons and neutrons were considered as a harmful radiation for organisms, while gamma rays, electrons, positrons have not been studied from that point.

Common to all works, 3 main factors inherent to an SN are considered to be the most powerful to induce biological effects on Earth. They are:

- gamma rays,
- X-rays,
- cosmic rays.

In the present study, we consider cosmic rays as the most powerful factor that affects the radiation background on Earth after a nearby SN explosion, and we investigate the possibilities of mutations in organisms and their possible mass extinctions.

## 2  Methods

We consider the CR fluxes $J(E,t)$ at the forward shock taken from the results of numerical simulations of the particle shock acceleration performed by Zirakashvili and Ptuskin (2012) for the SN type Ia, by Telezhinsky *et al.* (2013) for the SN type IIP and Caprioli (2011). Also, for comparison we used the CR spectra of Melott *et al.* (2017) with corrections from Thomas and Yelland (2023) as demonstrated in the following expressions:

$$n(E,L,t) = \frac{Q(E)}{(4\pi D(E)t)^{\frac{3}{2}}} \exp\left(-\frac{L^2}{4D(E)t}\right), \tag{1}$$

$$Q(E) \sim \left(\frac{E}{1\text{ GeV}}\right)^{-\gamma} \exp\left(-\frac{E}{1\text{ PeV}}\right), \tag{2}$$

where $D(E) = D_0 \left(\frac{E}{1\text{ GeV}}\right)^{\frac{1}{3}}$ is a diffusion coefficient, $D_0 = 2 \cdot 10^{28}$ cm$^2$/s, $\gamma = 2.2$, $L$ is the distance between Earth and an SN; the full CR energy taken here is $E_{CR} = 2.5 \cdot 10^{50}$ erg. It is typical for CR spectra, accelerated by Fermi I and II mechanisms, to be approximated by the power law in the energy range of $1\text{ GeV} \leq E \leq 10\text{ TeV}$, where the most of CR energy is contained.

All considered CR spectra are illustrated in Fig. 1, where triangles, squares, circles, and pentagons stand for different $L$. For the panel A, they are $L = 6.5$ pc, $L = 17$ pc, and $L = 42$ pc illustrated by triangles, squares, and circles, respectively. For the panel B, similarly, they are $L = 2$ pc, $L = 4$ pc, and $L = 8.4$ pc. For the panel C, they are $L = 15$ pc, $L = 26$ pc, $L = 41.7$ pc correspondingly, and $L = 56.4$ pc imaged by pentagons. And for the panel D, they are $L = 5$ pc, $L = 10$ pc, and $L = 20$ pc. Curves with stars represent the modern Local Interstellar Spectrum of CRs (LIS) (Usoskin *et al.*, 2005). We evaluated the time of the increase in the radiation background as the period during which the dose rate from the radiation background exceeded 10 rad/year. These characteristic times are presented in Table 1. Different acceleration and diffusion models and the type of the SN explosion impact the index $\gamma$ of the CR spectrum. Therefore, we performed the calculations with CR spectra by changing parameter $\gamma$. The CR spectra of kind (1) with $L = 5$ pc and $\gamma = 2.0$, $\gamma = 2.2$, $\gamma = 2.4$, $\gamma = 2.6$, and also for $\gamma = 1.7$ (as evidenced by Frolov *et al.* (2022)) are illustrated in Fig. 2.

Heliospheric modulation suppresses the low energy component of CR flux. If one were to suppose that the heliosphere is spherically symmetric with the homogeneous cosmic ray distribution in the Solar system, then the heliospheric modulation of CR spectra can be expressed as follows (Gleeson and Axford, 1968):

$$J_{mod}(E, \Phi, t) = J(E + \Phi, t) \frac{E(E + 2E_0)}{(E + \Phi)(E + \Phi + 2E_0)}, \quad (3)$$

where $E, E_0$ are the particle total energy, and its rest energy respectively, $\Phi = \frac{Ze\varphi}{A}$, is a modulation parameter for particles with the charge number $Z$ and mass number $A$, $\varphi$ is a modulation potential, and $e$ is the elementary charge. $J(E + \Phi, t)$ is the spectrum outside the heliosphere. For the SN event we have used both modulated ($\varphi = 360$ MV) and unmodulated ($\varphi = 0$ MV) particle spectra. The value $\varphi = 360$ MV is close to the average level of the Solar activity for the last 20 ky (Frolov *et al.*, 2022), while $\varphi = 0$ MV was used for the case of possible compression of the heliosphere due to the SN shockwave (Fry *et al.*, 2015; Fields *et al.*, 2008). For the considered CR spectra, the heliospheric modulation effects lead to $0.5 - 5$ % correction to unmodulated results.

The Earth's magnetic field leads to the energy (rigidity) cutoff of the CR spectra, where the rigidity of a particle is determined as:

$$R = \frac{pc}{Ze} = \frac{1}{Ze}\sqrt{E^2 + 2EE_0}. \quad (4)$$

Thus, all particles that have a rigidity lower than the cutoff rigidity $R_c = \mu R_0 \cos^4 \lambda$ at the current magnetic latitude $\lambda$ do not enter Earth's atmosphere. Here, $R_0 = 14.9$ GV, the cutoff rigidity at the geomagnetic equator, and $\mu$ -, the value of geomagnetic dipole moment in units of the current dipole magnetic moment of Earth. Here, we introduce the step function, which selects only particles with higher rigidity than a corresponding geomagnetic cutoff rigidity:

$$\Theta(E, \lambda) = \begin{cases} 1, & R(E) \geq R_c^\lambda, \\ 0, & R(E) < R_c^\lambda, \end{cases} \quad (5)$$

which is used in further calculations.

## 2.1 External ionization

Using Geant4 code, we carried out the nuclear cascade simulation in the atmosphere as triggered by ionizing radiation (Agostinelli *et al.*, 2003). Since any organisms consist of water by 60-70%, we simulated the energy absorption by mixed layers of the ocean assuming the same absorption mechanism in organisms. As a result of the modelling, we have the yield function for the external ionization $Y_{EI}(E, h)$, which is equal to the absorbed energy in the layer between $h$ and $h + \Delta h$, where $\Delta h = 5$ cm, per one CR proton with energy $E$. In our simulations, we considered only protons in CRs with energies 0.3 GeV $\leq E \leq$ 10 PeV. The absorbed dose rate of the external ionization is then

$$D_{EI}(h, \lambda, t) = \pi \int J_{mod}(E, \Phi, t) \cdot Y_{EI}(E, h) \cdot \Theta(E, \lambda) \, dE. \tag{6}$$

To obtain a biologically equivalent dose rate $D_{\text{bio}}$, a mean radiation weighting factor $\bar{\eta}$, depending on the particle kind and its energy, should be introduced for the considered CR spectra:

$$D_{\text{bio}} = \bar{\eta} D_{EI}. \tag{7}$$

The radiation weighting factors for different species of particles are given in Table 2 (ICRP, 2007). A mean radiation weighting factor can be obtained as

$$\bar{\eta} = \sum_i \eta_i F_i, \tag{8}$$

where $F_i$ is a fraction of $i$-th kind of particles by their quantity in the cascade.

## 2.2 Radiocarbon decay

Besides the external ionization by CRs, there is also an internal irradiation increase of organisms because of additional production of cosmogenic radionuclides in the atmosphere and their subsequent absorption. From the point of biological impact, the most effective isotope is $^{14}$C, because all organic molecules consist of carbon. Therefore, radiocarbon decay makes a significant contribution to the internal irradiation and, thus, potential DNA damage in cells. That raises the rate of mutations and the probability of extinctions if the $^{14}$C/$^{12}$C isotopic ratio is high enough. Moreover, the radiocarbon production rate is the highest among other radionuclides that results from the exothermic nature of its production reaction: $^{14}$N + n → $^{14}$C + p.

Using the results of our simulation in Geant4 code, we also calculated the radiocarbon yield function, $Y_{^{14}C}(E)$. This represents the number of $^{14}$C atoms generated in the atmosphere by one CR proton or $\alpha$-particle with the same range of kinetic energy $E$. After the production, radiocarbon atoms are rapidly oxidized to $^{14}CO_2$, which involves $^{14}$C in the global carbon cycle. Then $^{14}CO_2$ molecules spread across the atmosphere with the horizontal characteristic mixing time from several weeks to months and a vertical time of 1-2 years, and they are distributed to other reservoirs (mixed layers, biosphere, and deep ocean). Since the horizontal mixing time of radiocarbon is much less than its half-life time, we may take an average of $\Theta(E, \lambda)$ over the solid angle.

$$S(E) = \frac{1}{4\pi} \int \Theta(E, \lambda) d\Omega = \frac{1}{2} \int_{-\frac{\pi}{2}}^{\frac{\pi}{2}} \Theta(E, \lambda) \cos \lambda \, d\lambda = \text{Re}\left(1 - \sqrt{1 - \sqrt{\frac{R}{\mu R_0}}}\right), \tag{9}$$

where Re is the real part of the expression. The $^{14}C$ total production rate in the atmosphere is

$$Q(t) = \pi \int J_{\mathrm{mod}}(E, \Phi, t) \cdot Y_{^{14}C}(E) \cdot S(E) dE. \qquad (10)$$

The simplified model of the $^{14}C$ cycle, containing only 4 reservoirs (atmosphere, biosphere, mixed layers, and deep ocean) is given by the system of ordinary differential equations:

$$\frac{dN_i}{dt} = Q_i(t) - \lambda_d N_i - \sum_{j, j \neq i} \lambda_{ij} N_i + \sum_{j, j \neq i} \lambda_{ji} N_j, \qquad (11)$$

where $N_i$ is the radiocarbon concentration in the $i$-th reservoir, $\lambda_d$ is a radiocarbon decay probability, and $\lambda_{ij}$ is the transition probability for radiocarbon atoms from the $i$-th to the $j$-th reservoir.

Thus, we model two main ways of an increase in the irradiation of organisms on Earth in this work:

- the ionization by cascade particles (external irradiation),
- the radiocarbon decay inside organisms (internal irradiation).

## 3 Results

### 3.1 The effect of the external ionization

The average dose rates on Earth contributed by the external irradiation as a function of water depth are presented in Fig. 3. Curves with triangles, squares, and circles represent the external irradiation at the period, when the SN shock crosses the Solar System, for different distances $L$. Panel A illustrates the dose rates realized by the spectra from the work of Zirakashvili and Ptuskin (2012). In this panel, triangles stand for $L = 6.5$ pc, squares for $L = 17$ pc, and circles for $L = 42$ pc. In panel B, the dose rates are realized by the spectra from the work of Telezhinsky *et al*. (2013). Triangles - $L = 2$ pc, squares - $L = 4$ pc, and circles - $L = 8.4$ pc). Panel C represents the dose rates realized by spectra from Caprioli (2011). Here, triangles - $L = 15$ pc, squares - $L = 26$ pc, circles - $L = 41.7$ pc, and pentagons - $L = 56.4$ pc.

In panel D, the results of calculations with the spectra given by the equation (1) are also illustrated (triangles - $L = 5$ pc, squares - $L = 10$ pc, and circles - $L = 20$ pc). Curves with stars demonstrate the irradiation by the current LIS flux. Horizontal dashed lines represent the modern radiation background from all terrestrial sources, which is $\approx 0.24$ rad/year (UNSCEAR, 2008). The dose rates realized by the spectra of kind (1) for $L = 5$ pc and time $t = L^2/6D_0$, corresponding to the maximum level of CR intensity. Different $\gamma$ are demonstrated in Fig. 4.

The dose rate values for $L = 50$ pc and $t = 100, 300, 1000, 3000, 10000$ years after the arrival of the first photons from the SN are illustrated in Fig. 5. Fig. 6 demonstrates the fractions of cascade particles by their number and energy at sea level. Using these data, one can obtain the mean radiation weighting factor $\overline{\eta}$. For the CR spectra (1) with $\gamma = 2.2 - 2.7$ it is $\overline{\eta} = 1.17 - 1.19$.

## 3.2 The effect of the radiocarbon decay

The $^{14}$C production rate in the atmosphere and the dose rates resulting from $^{14}$C decay in the biosphere are presented in Fig. 7 as a function of time for $L = 6.5$ pc (panels A, D), $L = 17$ pc (panels B, E), and $L = 42$ pc (panels C, F) with the CR spectra from Zirakashvili and Ptuskin (2012). To take the possible geomagnetic reversal into account, we carried out the calculations with different values of $\mu$. The value $\mu = 0.1$ corresponds to the reversal period (the dash-dotted curves) and $\mu = 1$ - to the current magnetic field of Earth (the solid curves). The present level of $^{14}$C production rate is estimated to be 1.7 atoms/(cm$^2 \cdot$ s) (dotted line in the panel C). Other considered CR spectra lead to the dose rates contributed by $^{14}$C decay below the current level of the radiation, and therefore we do not consider them.

The absorbed dose can be accumulated by organisms for the years up to their whole lifetime. The instantaneous doses that result in 50% probability of lethal outcome ($LD_{50}$) according to ICRP (2008) are demonstrated in Table 3. There is a proportion between instantaneous lethal doses for organisms and threshold annual doses for depression in the immune system ($LD_{th}$), which are supposed in this work to form the radiation background accroding to ICRP (2007, 2008) and Akleyev *et al.* (1999) for a human $LD_{50} = 600$ rad and $LD_{th} = 30$ rad/year. We consider the next lower and upper bounds on the lethal dose rates: $D_{le}^{max} = 600$ rad/year and $D_{le}^{min} = 30$ rad/year. Figure 8 shows the maximal values of $L_{le}$ for all considered CR spectra, where the mentioned effects on organisms are possible.

## 4. Discussion

The irradiation of organisms can be classified as an external irradiation and an internal one. The external irradiation makes the greatest contribution to the absorbed dose rate for all kinds of organisms. Fig. 6 demonstrates that the nuclear cascade in the atmosphere is mainly driven by $\gamma$-radiation, muons, electrons, positrons, and neutrons. As for the energy fractions of particles in the cascade, we found that the most energy, which is about 95%, is contained in muons, while only 2% of energy are in photons and $\approx 1.5$ % in neutrons.

Our estimations of the absorbed dose rates in the framework of pure diffusion model (Figs. 4, 5) are lower than those obtained by Melott *et al.* (2019) and Thomas and Yelland (2023) within the same model. This mismatch could be related to the different usage of numerical codes. Our calculations of the yield functions are driven only by the Geant4 code, while Melott *et al*. (2019) and Thomas and Yelland (2023) used Geant4 and CORSICA together. Therefore, similar calculations with the yield functions obtained in previous works but with the same CR spectra would result in higher absorbed dose rates. There is also the well-known problem of an instantaneous propagation, which appears in the solutions (see, for example, (1)) of parabolic equations. This paradox is resolved in the relativistic diffusion equation (Cattaneo, 1958; Chester, 1963). However, for the sake of simplicity, in the works of Melott *et al.* (2017) and Thomas and Yelland (2023), the function (1) is considered at the moments of time from the arrival of the first photons. In other words, they cut down the solution of the parabolic equation at distances $r > ct$, where $c$ is the speed of light, eliminating the effects of instantaneous propagation of particles. This simplest approximation is possible for the estimation of the effects they considered in their works. However, we claim that it is important to consider the CR acceleration and transport by an SN shock. In this model, the CR flux has the steep jump at the shock, unlike the diffusive model, where the CR flux due to the absence of shockwave structure is smooth everywhere as a function of coordinate.

The effects considered in the present study have occured on Earth at least twice for the last 10 My, as evidenced by the measurements of the $^{60}$Fe concentration in deep sea crusts. The latest such event, dated 2.6 Ma, could be a trigger for the extinction of organisms at the PP boundary including the impact of the possible geomagnetic reversal at that time. According to $^{60}$Fe detections (Fry *et al*., 2015), the core-collapse SN (SN II type) is most likely to occur at 2.6 Ma. Also, the average rate of SN II explosions in our Galaxy is about 6 times higher than the rate of SN Ia explosions (Tsujimoto *et al*., 1995). Thus, the results of Telezhinsky *et al*. (2013), where the acceleration by the SN IIP shockwave is studied, is more applicable to that event. However, the variation of other parameters of the acceleration model ($E_{SN}$, $L$, interstellar medium density etc.) can more strongly affect the CR spectra than SN types. Therefore, we do not compare the effects caused by different kinds of SNe.

The characteristic time of the CR flux increase near Earth is much longer than the typical organism lifetimes or comparable to them, as in the case considered by Telezhinsky *et al*. (2013). Thus, the external irradiation of organisms exceeds the modern radiation background by several orders of magnitude and can result in harmful effects for the terrestrial biosphere and for the mixed ocean layers fauna.

The internal irradiation is mostly realized by the radiocarbon decay. The effective absorbed dose rate that results from $^{14}$C decay inside organisms can be 10-100-fold greater than that from the external ionization (Busby, 2013). This is because almost all organics are comprised of $\sim$ 25% carbon. Therefore, $^{14}$C decay destroys organic molecules including DNA, which enhances the mutational and lethal effects. The maximal dose rates contributed by the radiocarbon decay at $L \lesssim 10$ pc are comparable with critical dose rate for the staff of atomic power plants, which is estimated to be 2 rad/year (ICRP, 2007). Such dose rates are insufficient to make mass extinctions possible, but the rate of mutations and the risk of getting cancer still take place, especially in the deep ocean at depths $\geq 100$ m, where external ionization can not occur. At $L > 20$ pc, the radiocarbon decay contribution is not significant. Besides the radiocarbon, there is also increased production of cosmogenic tritium in the atmosphere (though 4-6 times lower than $^{14}$C). $^3$H decay ($T_{1/2}(^3H) = 12.32$ years) can also produce DNA damage. Tritium atoms generated in the atmosphere are rapidly oxidized to $H_2O$ molecules, which accumulate in the Ocean and different water reservoirs on the ground. As a result, $^3$H amount in water absorbed by different organisms is mainly uncertain and probably low. Therefore, the modelling of the biological impact of $^3$H is difficult and is beyond the scope of this work.

As for the possibility of mutations, it is clear that their average rate always becomes higher with an increase in CR flux and, therefore, in the radiation background on Earth. However, it is not easy to do a strict analysis of mutation rates in organisms given that it requires many additional investigations. Namely, different $E_{SN}$, $L$, interstellar medium density, and conversion efficiency of SN kinetic energy into CRs, which could lead to significantly different biological effects.

It should be noted that CRs can affect the terrestrial and marine biosphere not only by way of external irradiation and the radiocarbon decay; there are also such factors as atmospheric ozone depletion, X-rays, UV, and optical emission from an SN, which is beyond the scope of this work. The effect of $O_3$ depletion was evaluated earlier by Melott *et al*. (2017), Brunton *et al*. (2023), and Thomas and Yelland (2023). The latter work claims the depletion of $\geq 40\%$ of the atmospheric ozone from a "standard" SN, which takes place at $L = 20$ pc. Therefore, this effect can make an additional contribution to the considered CR influence on the biosphere.

# 5 Conclusion

We found that cosmic rays from nearby SNe are able to trigger mutations and mass extinctions of terrestrial organisms. The maximal value of the distance between the SN and Earth, where the lethal effects are possible, is obtained with the CR spectra from the work of Zirakashvili and Ptuskin (2012) with $E_{SN} = 10^{51}$ erg, which is $L_{le} = 18$ pc. It is consistent with the "lethality distance" of 20 pc from the work of Thomas and Yelland (2023), based on ozone depletion by nearby SN CRs, while the estimation of $L_{le}$ by Brunton *et al*. (2023), based on X-ray luminous supernova, is in the range of $20 - 50$ pc. For other considered spectra from the works of Telezhinsky *et al*. (2013), Caprioli (2011), and Melott *et al*. (2017), $L_{le} = 3 - 8$ pc, which is of the order of $L_{le} = 8$ pc from the ozone depletion effects, as obtained by Gehrels *et al*. (2003).

The contribution of the radiocarbon decay to the radiation background is much lower than that of the external irradiation and is not sufficient to make mass extinctions possible. The highest biological effect of the external irradiation is found to occur inside the first 100 m of a water depth (mixed ocean layers). There are no noticeable changes in the radiation background in the deep ocean. However, $^{14}$C makes the contribution to the internal irradiation of organisms, since it spreads over the whole biosphere and the ocean, and the dose rates of internal irradiation can be $10 - 100$ times higher than the characteristic dose rates by the rest of the radiation background.

Table 1. Characteristic duration times $\tau$ of the CR flux increase near Earth

|  | $L$, pc | $\tau$, years |
|---|---|---|
| Zirakashvili and Ptuskin (2012) | 6.5 | 1000 |
|  | 17 | $> 20000$ |
| Telezhinsky *et al.* (2013) | 2 | 130 |
|  | 4 | 22 |
| CR spectra of kind (1) | 5 | 3700 |

Table 2: Radiation weighting factors $\eta$ for different particle species generated in an atmospheric nuclear cascade taken from ICRP (2007)

| Particle | Energy range | $\eta(E, \text{MeV})$ |
|---|---|---|
| Neutron | $< 1$ MeV | $2.5 + 18.2 \exp\left(-\dfrac{\ln^2(E)}{6}\right)$ |
|  | $1 \text{ MeV} - 50 \text{ MeV}$ | $5.0 + 17.0 \exp\left(-\dfrac{\ln^2(2E)}{6}\right)$ |
|  | $> 50$ MeV | $2.5 + 3.25 \exp\left(-\dfrac{\ln^2(0.04 \cdot E)}{6}\right)$ |
| Proton | All | 2 |
| $e^-, e^+, \mu^-, \mu^+, \gamma$ | All | 1 |
| $\pi^-, \pi^+$ | All | 2 |

Table 3: $LD_{50}$ doses for animals according to ICRP (2008)

|  | $LD_{50}$, rad |
|---|---|
| Large mammals | $120 - 800$ |
| Small mammals | $600 - 1000$ |
| Birds | $500 - 1200$ |
| Fishes | $30 - 1900$ |

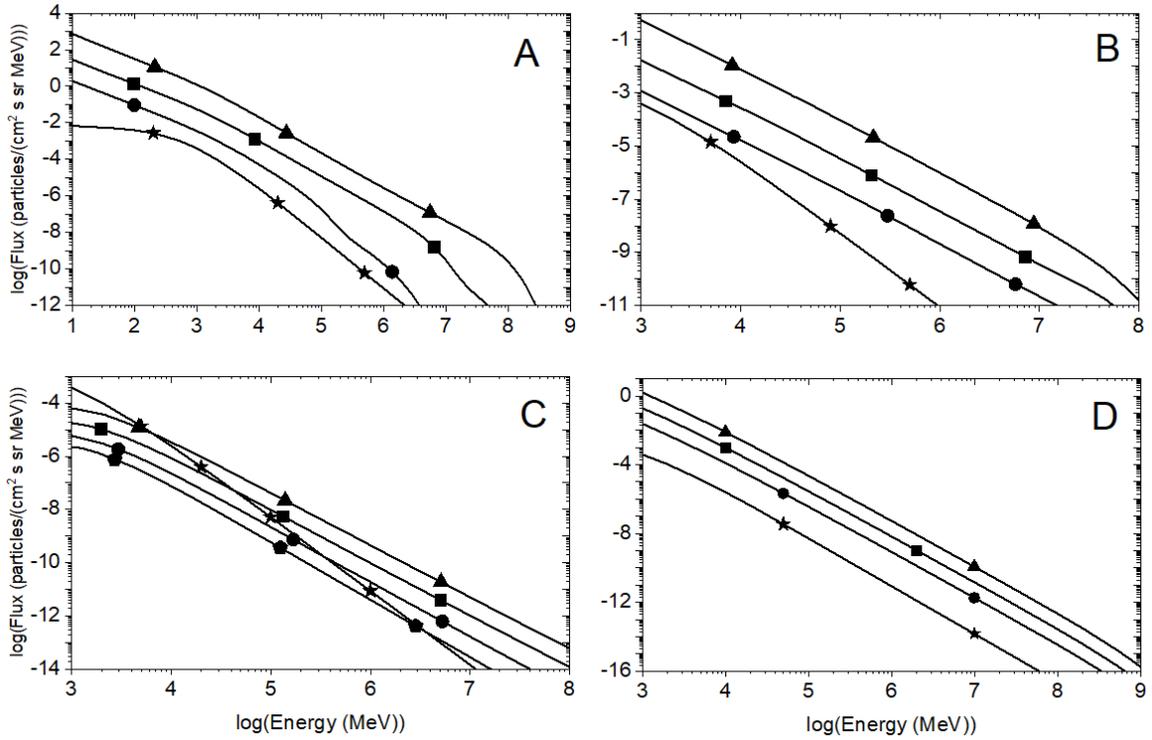

Figure 1: Cosmic ray spectra at the forward shock taken from Zirakashvili and Ptuskin (2012) (A), Telezhinsky *et al.* (2013) (B), Caprioli (2011) (C) and the CR spectra of kind (1) from Melott *et al.* (2017) (D). Triangles stand for $L = 6.5$ pc, $L = 2$ pc, $L = 15$ pc and $L = 5$ pc for the panels A, B, C, D respectively. Squares represent $L = 17$ pc, $L = 4$ pc, $L = 26$ pc and $L = 10$ pc for the same panels. Circles stand for $L = 42$ pc, $L = 8.4$ pc, $L = 41.7$ pc and $L = 20$ pc. Pentagons in the panel C represent $L = 56.4$ pc, and stars in all panels stand for the modern Local Interstellar Spectrum of CRs (Usoskin *et al.*, 2005)

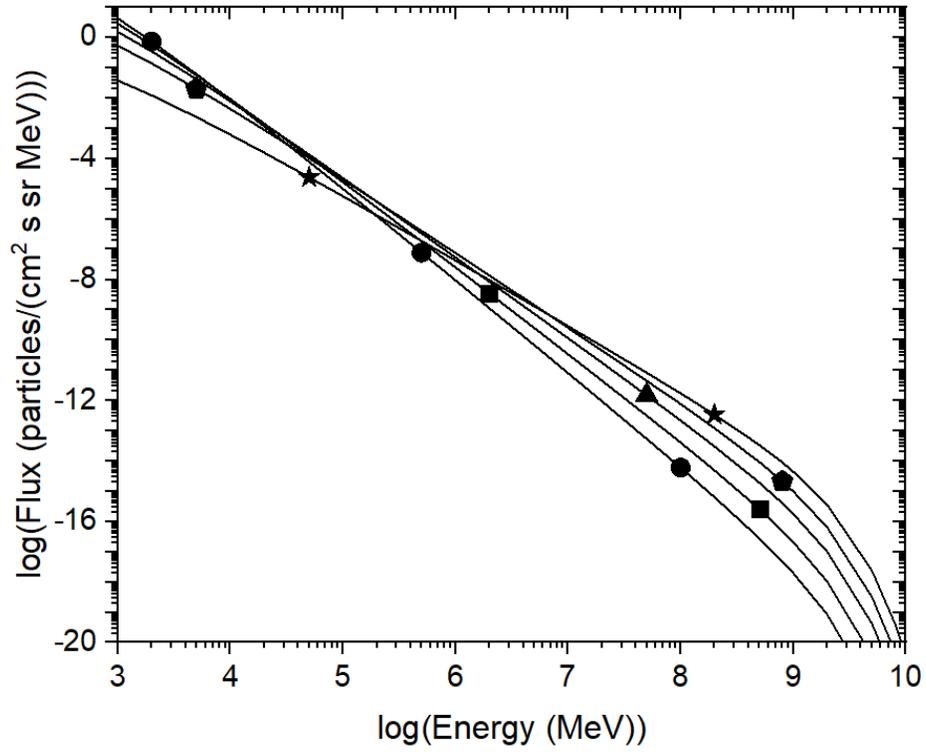

Figure 2: CR spectra (1) corresponding to $L = 5$ pc and $\gamma = 1.7$ (stars), $\gamma = 2.0$ (pentagons), $\gamma = 2.2$ (triangles), $\gamma = 2.4$ (squares) and $\gamma = 2.6$ (circles)

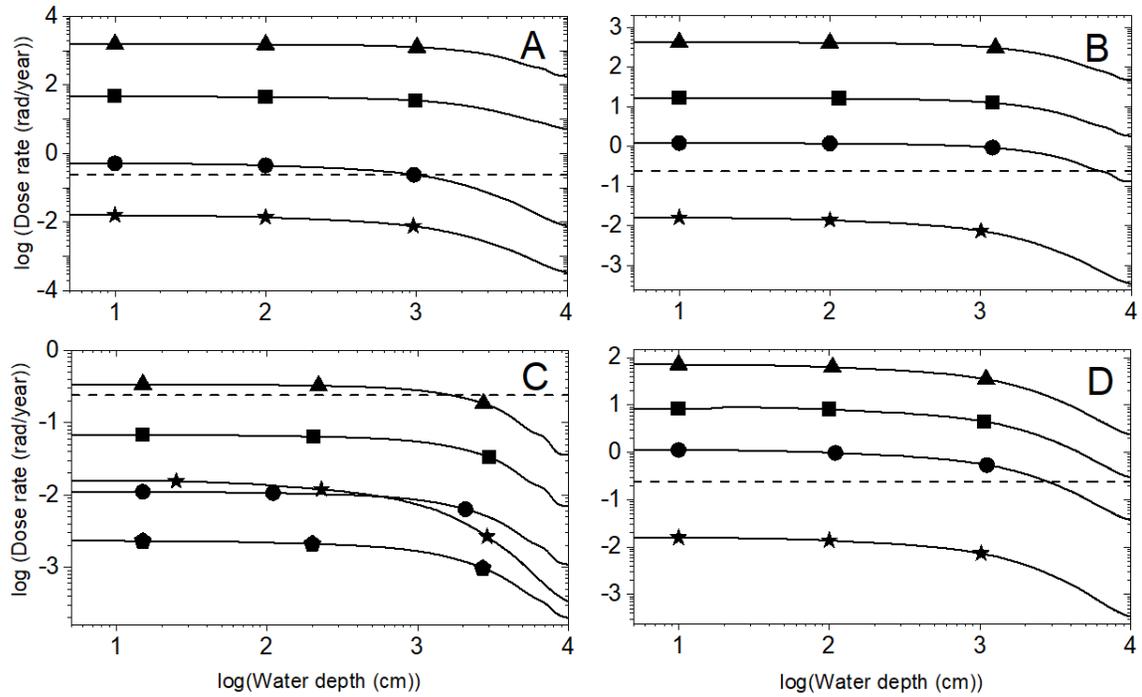

Figure 3: The average absorbed dose rates of external irradiation as functions of the water depth (see Fig. 1 for the description). Horizontal dashed lines represent the modern radiation background from all terrestrial sources (UNSCEAR, 2008)

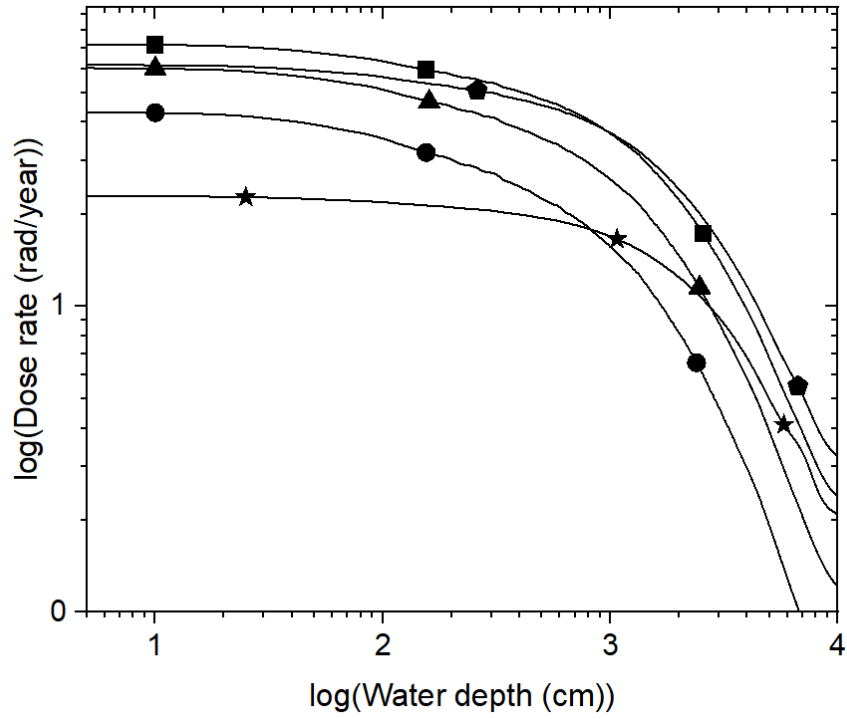

Figure 4: Dose rates obtained using the spectra (1) as functions of the water depth at $L = 5$ pc and $\gamma = 1.7$ (stars), $\gamma = 2.0$ (pentagons), $\gamma = 2.2$ (squares), $\gamma = 2.4$ (triangles) and $\gamma = 2.6$ (circles)

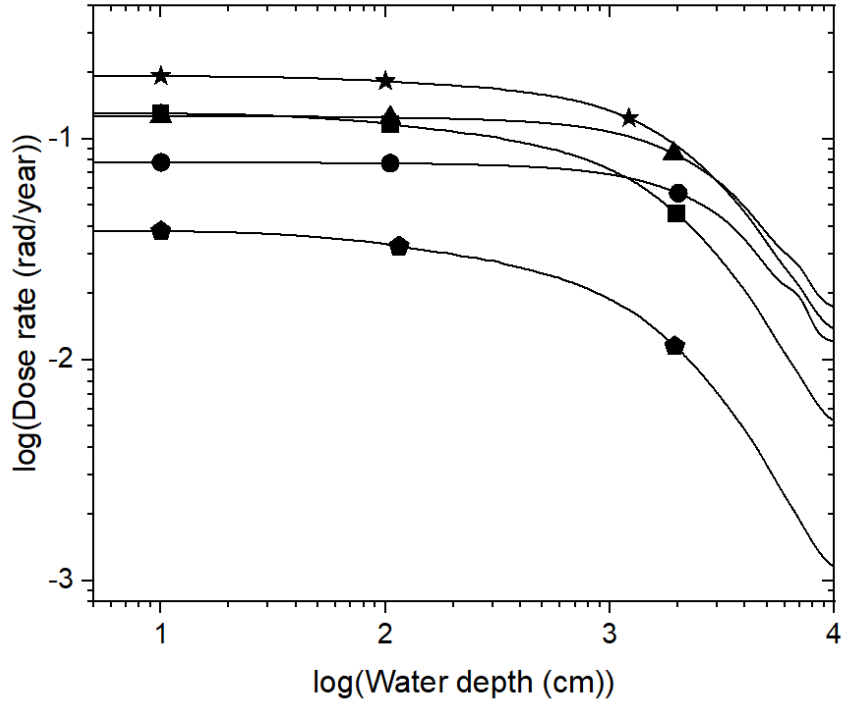

Figure 5: Dose rates obtained using the spectra (1) as functions of water depth at $L = 50$ pc, $\gamma = 2.2$ and $t = 100$ years (circles), $t = 300$ years (triangles), $t = 1000$ years (stars), $t = 3000$ years (squares) and $t = 10000$ years (pentagons)

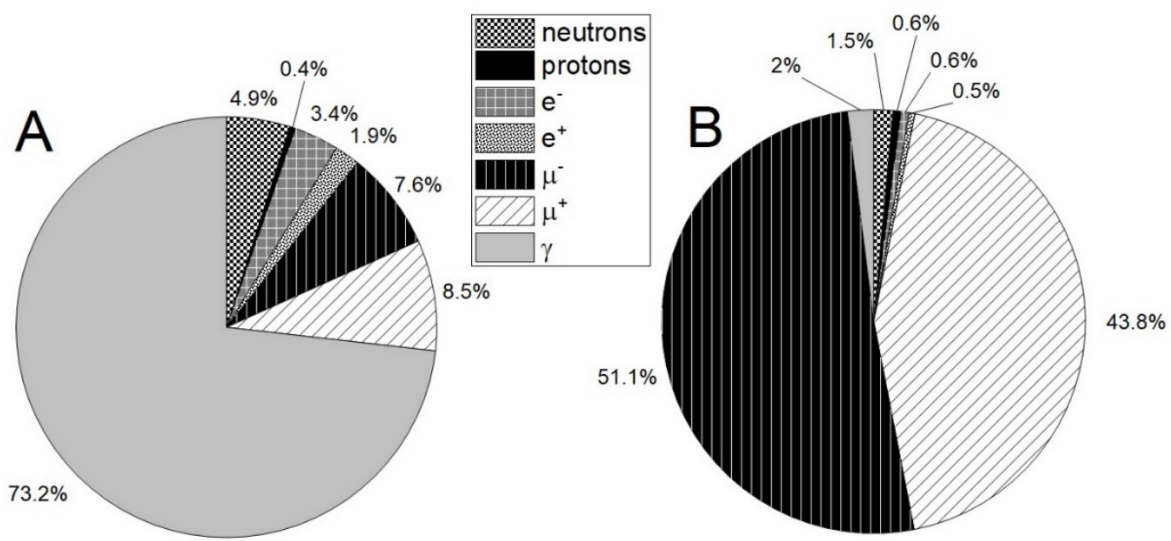

Figure 6: The fractions of cascade particles by their quantity (A) and by their energy (B) at sea level

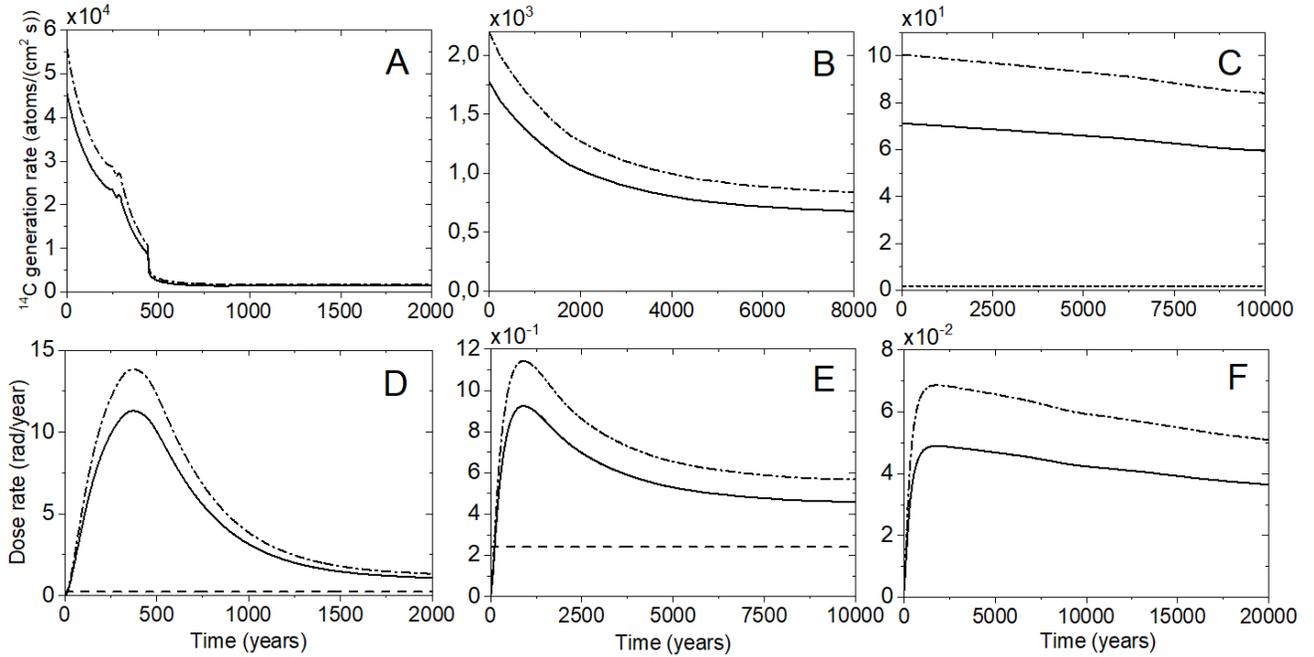

Figure 7: The $^{14}C$ production rates in the atmosphere (panels A, B, C) and absorbed dose rates in the biosphere (panels D, E, F) resulted from $^{14}C$ decay and calculated using the CR spectra from Zirakashvili and Ptuskin (2012) as functions of time for $L = 6.5$ pc (A, D), $L = 17$ pc (B, E), $L = 42$ pc (C, F). The moment $t = 0$ corresponds to the moment of the maximum of the CR intensity, when the SN shock crosses the Solar System. Dotted horizontal line in the panel C and dashed horizontal lines in the panels D and E represent the current $^{14}C$ production rate in the atmosphere ($\approx$ 1.7 atoms/(cm$^2 \cdot$ s)) and the current radiation background from all terrestrial sources (0.24 rad/year) respectively. The case $\mu = 0.1$ is illustrated by the dash-dotted curves, and the case $\mu = 1$ is shown by the solid curves

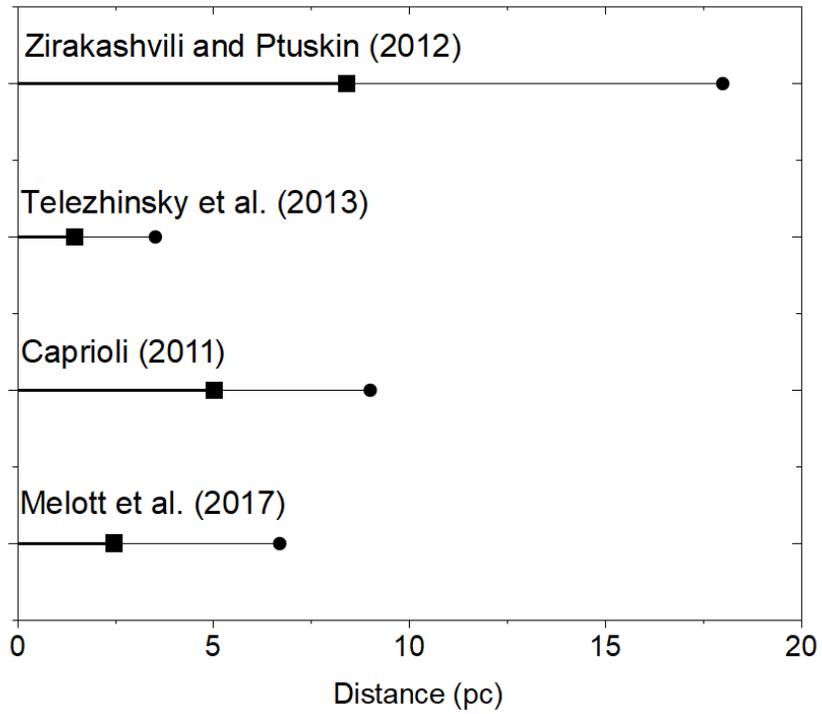

Figure 8: The maximal distances $L_{le}$ between Earth and an SN, where lethal effects are possible. Squares and circles represent points at which the maximal dose rate at the surface of Earth is 600 rad/year and 30 rad/year respectively